\documentclass[aps,prl,groupedaddress,twocolumn,10pt,showpacs]{revtex4}
\usepackage[pdftex]{graphicx}
\usepackage{amsmath,amssymb,color,subfigure,yllmath}
\providecommand{\myheading}[1]{\textbf{#1}}
\providecommand{\omegados}{{\omega_\text{dos}}}
\providecommand{\Deltaop}{{\Delta_\text{op}}}

\begin{document}
\title{Single and two-particle energy gaps across the disorder-driven superconductor-insulator transition}
\author{Karim Bouadim}
\author{Yen Lee Loh}
\author{Mohit Randeria}
\author{Nandini Trivedi}
\affiliation{Department of Physics, The Ohio State University, Columbus, OH  43210, USA}
\begin{abstract}
The competition between superconductivity and localization raises profound questions in condensed matter physics. In spite of decades of research, the mechanism of the superconductor-insulator transition (SIT) and the nature of the insulator are not understood. We use quantum Monte Carlo simulations that treat, on an equal footing, inhomogeneous amplitude variations and phase fluctuations, a major advance over previous theories. We gain new microscopic insights and make testable predictions for local spectroscopic probes. The energy gap in the density of states survives across the transition, but coherence peaks exist only in the superconductor.  A characteristic pseudogap persists above the critical disorder and critical temperature, in contrast to conventional theories. Surprisingly, the insulator has a two-particle gap scale that vanishes at the SIT, despite a robust single-particle gap.
\end{abstract}
\maketitle

Attractive interactions between electrons lead to superconductivity, 
a spectacular example of \emph{long range order} in physics, 
while disorder leads to \emph{localization} of electronic states.
One of the most fascinating examples of the interplay between the
effects of interactions and localization is
the destruction of superconductivity in thin films with increasing disorder 
and the resulting superconductor-to-insulator transition (SIT)~\cite{goldman-markovic_physicstoday1998,
gantmakher2010, haviland1989, hebard1990, shahar1992, adams2004, steiner2005, valles_nanohc_science2007,sachdev_qpt}.

It was recognized several decades ago that $s$-wave superconductivity (SC) 
is remarkably robust against weak disorder~\cite{anderson1959,abrikosovgorkov1959}.
It was later argued~\cite{ma1985} that SC can survive even when disorder localizes the single-particle states.
Thus the  superconductor-to-insulator transition must occur in a strong disorder regime 
that is difficult to treat theoretically in an interacting system.
Critical phenomena at the SIT have been described in terms of 
disordered bosons~\cite{fisher-grinstein-girvin_prl1990}, which model fermion pairs 
and describe phase fluctuations of the SC order parameter.
A more microscopic description must necessarily start with the fermionic degrees of freedom.
A Bogoliubov-de Gennes (BdG) treatment of attractive electrons in a random potential shows that
the SC pairing amplitude becomes spatially inhomogeneous with strong disorder~\cite{ghosal1998,ghosal2001,feigelman_prl2007}.
This leads to a robust energy gap and a large suppression of the superfluid density~\cite{ghosal1998,ghosal2001}. 
However, the phase fluctuations 
ultimately responsible for the SIT are beyond the BdG approach 
and are treated in an approximate manner~\cite{ghosal1998,ghosal2001,dubi_nature2007}.

In this paper we make a major advance using quantum Monte Carlo (QMC) simulations 
on a fermionic model, which  include thermal and quantum fluctuations of the SC phase 
\emph{and} the spatially inhomogeneous amplitude on an equal footing.
While confirming the bosonic mechanism for the SIT, our work also gives new insights into the experimentally observable \emph{electronic} spectral functions.  Our results provide us with a detailed description of the phases,
the transition, and the quantum critical region at finite temperature.  
 
Our main results are as follows:
\\
(1) \emph{Single-particle gap:}
At $T=0$ the gap in the single-particle density of states (DOS) survives through the SIT,
so that one goes from a gapped superconductor to a gapped insulator. 
Although the local gap extracted from the local density of states (LDOS) 
is highly inhomogeneous, it is nevertheless finite at every site.
\\
(2) \emph{Coherence peaks:}
These characteristic pile-ups in the DOS at the gap edges 
are directly correlated with superconducting order and vanish 
as the temperature is raised above $T_c$, 
or as the disorder is increased across the SIT. 
\\
(3) \emph{Pseudogap:}
Near the SIT, a pseudogap -- a suppression in the low-energy DOS -- 
persists well above the superconducting $T_c$ up 
to a crossover temperature scale $T^\ast$, in marked deviation from BCS theory.
This disorder-driven pseudogap also exists at finite temperatures in the insulating state
and grows with disorder.  
\\
(4) \emph{Two-particle gap:}
There is a characteristic energy scale $\omega_{\rm pair}$ to insert a pair in the insulator 
that collapses upon approaching the SIT from the insulating side. 
In addition the two-particle spectral function may also have 
very small spectral weight at low energies coming from
rare regions.

Our predictions for the local tunneling density of states and the dynamical pair susceptibility as a function of temperature and disorder have the potential to guide future experiments using 
scanning tunneling spectroscopy (STS)~\cite{sacepe2008,cren2000,lang2002,yazdani07} and other dynamical probes~\cite{crane_fluctuations_2007}. 

\myheading{Model and methods:}
To model the competition between superconductivity and localization that leads to the SIT in quench-condensed films with thicknesses less than the 
coherence length, we take the simplest lattice Hamiltonian that has an $s$-wave superconducting ground state in the absence of disorder ($V=0$)
and exhibits Anderson localization when the attractive interaction is turned off ($U=0$).
Thus, we study the two-dimensional attractive Hubbard model in a random potential:
	\begin{align}
	H&= -t\sum_{\langle \RRR \RRR' \rangle \sigma} 
			(\cdag_{\RRR\sigma} \cccc_{\RRR'\sigma} + \cdag_{\RRR'\sigma} \cccc_{\RRR\sigma})\nonumber \\
	&- \sum_{\RRR\sigma} (\mu-V_\RRR) n_{\RRR\sigma}
		-	|U|	\sum_\RRR n_{\RRR\uparrow} n_{\RRR\downarrow}.
	\label{Hamiltonian}
	\end{align}
with lattice sites $\RRR$ and $\RRR'$, spin indices $\sigma=\up$ or $\dn$, 
fermion creation and annihilation operators $\cdag_{\RRR\sigma}$ and $\cccc_{\RRR\sigma}$,  
number operators $n_{\RRR\sigma} = \cdag_{\RRR\sigma} \cccc_{\RRR\sigma}$,
hopping $t$ between neighboring sites $\langle \RRR\RRR' \rangle$,
and a chemical potential $\mu$ chosen such that the average density is $\mean{n} \neq 1$.
$V_\RRR$ is a random potential at each site drawn from the uniform distribution on $[-V,+V]$, 
and $\left| U \right|$ is the on-site attraction leading to $s$-wave SC.
We will measure all energies in units of $t$.

	\begin{figure}[!htb]
	\includegraphics[width=0.45\textwidth]{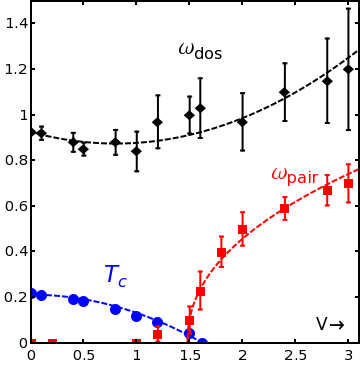}
	\caption{
	\label{PhaseDiag}
	\textbf{Energy and temperature scales across the superconductor-insulator transition (SIT).}
	The superconducting $T_c$ (blue dots) decreases to zero at the critical disorder strength $V_c$.
	The single-particle gap $\omegados$ (black diamonds),
	obtained from the DOS shown in Fig.~\ref{dos}, is large and finite in all states.
	The two-particle energy scale $\omega_{\rm pair}$ (red squares),
	obtained from the dynamical pair susceptibility shown in Fig.~\ref{PAIRFIG},
	is non-zero in the insulator but vanishes at the SIT.
	The dashed curves are guides to the eye; 
	extracting critical exponents requires finite-size scaling beyond the scope of this paper.
       The statistical error bars in all the figures are dominated by disorder averaging and not from the QMC.
	These results are obtained at fixed attraction $|U|=4$ and average density $\mean{n} \approx 0.87$ 
	on 10 disorder realizations on $8\times 8$ lattices.
	$\omega_{pair}$ and $\omegados$ are calculated at the lowest accessible temperature, 
	$T=0.1$.
	For specific parameter values, we have run extensive simulations 
	that average over	100 disorder realizations.   
	}
	\end{figure}

We use the determinantal QMC method~\cite{blankenbecler},
which is free of the fermion sign problem for the Hamiltonian \eqref{Hamiltonian}. 
We choose $\left| U \right| = 4$, so that the coherence length is within the system size,
and $\mean{n}=0.875$.
We have made extensive comparisons of the QMC results with self-consistent BdG calculations, 
which take into account only the spatial amplitude variations; see supplementary material.
These comparisons permit us to separate the effects of amplitude inhomogeneity and phase fluctuations. 

We compute frequency-dependent observables across the SIT for the first time.
The single-particle DOS, LDOS and the pair susceptibility are obtained using the maximum entropy method (MEM) for analytic continuation~\cite{jarrell,sandvik1998}.
We have verified that these results obey various sum rules to high precision, and that the MEM correctly reproduces the low-energy structure of test spectra as shown
in the supplementary material.
What gives us confidence is that our central results on the single- and two-particle gaps can be equally well
estimated directly from the exponential decay of the imaginary-time QMC data, without recourse to MEM. 

\begin{figure*}[!htb]
\includegraphics[width=\textwidth]{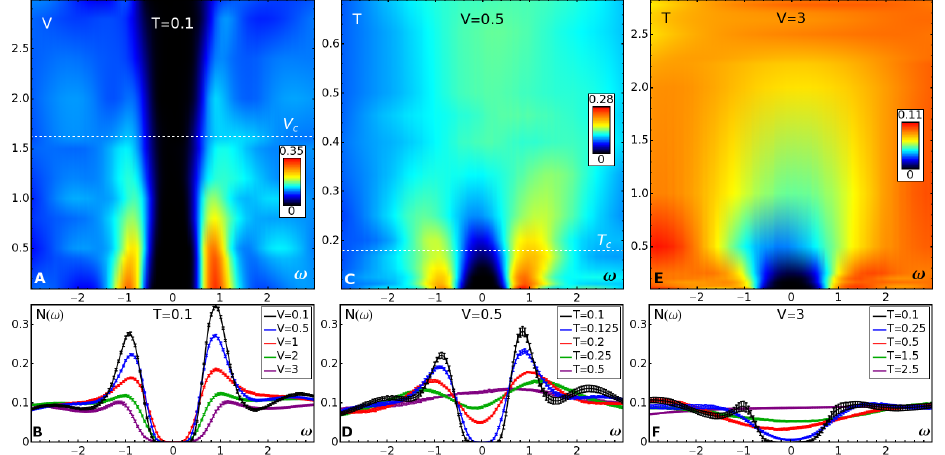}
\caption {
\label{dos}
\textbf{The single-particle DOS} $N(\omega)$  (upper panels) and representative spectra (lower panels) along three different cuts through the temperature-disorder plane.
Left panels (A,B): Disorder dependence of $N(\omega)$ at a fixed low temperature. 
A hard gap (black region) persists for all $V$ above and below the SIT ($V_c \approx 1.6$),
but the coherence peaks (red) exist only in the SC state and not in the insulator. 
Center panels (C,D): $T$-dependence of the $N(\omega)$ for the superconductor ($V< V_c$). 
The coherence peaks (red) visible in the SC state, vanish for $T\gtrsim T_c \approx 0.14$.
A {\em disorder-induced pseudogap}, with loss of low-energy spectral weight, 
persists well above $T_c$.
Right panels (E,F): $T$-dependence of $N(\omega)$ for the insulator ($V>V_c$).  The hard insulating gap 
at low $T$ evolves into a pseudogap at higher $T$. No coherence peaks are observed at any $T$.
All panels show data averaged over 10--100 disorder realizations.
}
\end{figure*}

\myheading{Phase diagram:} 
In Fig.~\ref{PhaseDiag} we summarize our key results for the disorder dependence of various
temperature and energy scales.
Since the finite temperature transition is expected to be in the Berezinskii-Kosterlitz-Thouless
universality class, we estimate the critical temperature $T_c$ 
from the superfluid density $\rho_s$, calculated from the transverse current correlator~\cite{scalapinowhitezhang1993,trivedi1996}.
We note that this procedure on finite systems provide an upper bound on the actual
$T_c$ in the thermodynamic limit.
As disorder strength $V$ increases, $T_c$ falls and finally vanishes at the critical
disorder $V_c$, which defines the SIT.
The single-particle energy gap $\omegados$ remains non-zero across the SIT, 
whereas the two-particle energy scale $\omega_{\rm pair}$ is finite in the insulator and goes to zero at the transition. 
These gap scales are extracted from the DOS and the dynamical pair susceptibility discussed 
in detail below.
Figure~\ref{PhaseDiag} can be interpreted as a phase diagram: $T_c$ is the superconducting transition temperature,
$\omega_{\rm pair}$ is a crossover scale between the insulator and the quantum critical region, and $\omegados$ is
related to the pseudogap crossover scale described below. 

\myheading{Single-particle spectra:}
We show in Fig.~\ref{dos} the disorder and temperature dependence of the DOS $N(\omega)$.
Panels (A,B) show the evolution with disorder at a very low temperature $T=0.1$.
The gap $\omegados$ \emph{clearly remains finite} in both superconducting and insulating states,
a counterintuitive observation that agrees qualitatively with BdG results~\cite{ghosal1998,ghosal2001}.
In contrast, the coherence peaks diminish with increasing $V$ and disappear near the SIT at $V_c \approx 1.6$.

Figures~\ref{dos}(C,D)  show the temperature evolution of $N(\omega)$ at weak disorder $V < V_c$.
Unlike in BCS theory, the hard SC gap does not close with increasing $T$.  
Instead, the coherence peaks gradually disappear as the temperature increases across $T_c$.	
Above $T_c$, the gap gradually fills up, with a pseudogap persisting well above $T_c$.

The temperature evolution of $N(\omega)$ at strong disorder $V > V_c$ is shown in Fig.~\ref{dos}(E,F).
Here the ground state is an insulator with a hard gap and little evidence for coherence peaks,
and the pseudogap persists up to an even higher temperature.

\begin{figure}[!htb]
	\includegraphics[width=0.42\textwidth]{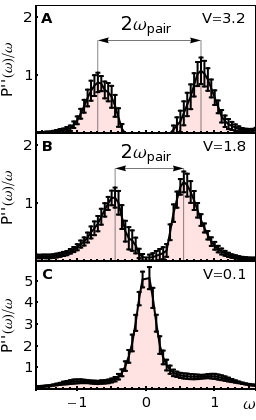}
	\caption{
	\label{PAIRFIG}
	\textbf{Imaginary part of the dynamical pair susceptibility} $P''(\omega) / \omega$ at $T=0.1t$,
	averaged over 10 disorder realizations at three disorder strengths.
	Error bars represent variations between disorder realizations.
	For $V<V_c$, there is a large peak at $\omega=0$, indicating zero energy cost to insert a pair into the SC.
	For $V>V_c$, there is a gap-like structure with an energy scale $\omega_{\rm pair}$,
	the typical energy required to insert a pair into the insulator which increases with $V$.
	}
\end{figure}

\myheading{Two-particle spectra:}
Given that we find an insulator with a single-particle gap, 
what is the energy scale that vanishes upon approaching the quantum critical point
from the insulating side? 
We propose that it is the typical energy for a \emph{two-particle excitation} in the insulator. 
To access this scale, we examine the pair susceptibility $P(\omega)$ 
obtained by analytical continuation of the correlation function 
$P(\tau) = \sum_\RRR \mean{\calT_\tau F (\RRR;\tau) F^\dag (\RRR;0)}$ 
where $F(\RRR,\tau)=  \cccc_{\RRR\dn} (\tau) \cccc_{\RRR\up} (\tau)$.  
Thus $P(\tau)$ is the amplitude for a pair created at a site $\RRR$ at $\tau=0$ to be found at the 
same site at a later time $\tau$. We find that in the insulating phase
$P(\tau)$ decays exponentially, which allows us to define $\omega_{pair}$, the characteristic energy scale for two-particle excitations.

In Fig.~\ref{PAIRFIG} we show the imaginary part of the pair susceptibility 
$P''(\omega)/\omega$ for three disorder strengths.
At weak disorder $P''(\omega)/\omega$ is very large at low $\omega$, 
whereas at strong disorder it has a clear two-peak structure with a characteristic energy scale $\omega_{\rm pair}$.
This dominant scale represents the typical energy required to insert a pair into the system. 
We find that $\omega_{\rm pair}$ collapses to zero at the SIT because there is no cost for inserting a pair into a condensate.

At sufficiently small energies our insulating state is similar to a Bose glass, in which rare regions~\cite{fisher1989} give 
rise to a very small but non-zero spectral weight in $P''(\omega)/\omega$ at low energies.
Such Griffiths-McCoy-Wu singularities can be very difficult to pin down
in numerical simulations and even in experiments. Nevertheless, we do indeed see
some signs of low-energy spectral weight in, e.g., Fig.~\ref{PAIRFIG}B.  
In this paper, however, we focus on the most salient features in 
$P''(\omega)/\omega$. These are the \emph{peaks} at $\pm\omega_{\rm pair}$,  
which imply that the \emph{typical} energy cost to insert a pair is finite.

\begin{figure*}[!htb]
	\includegraphics[width=0.8\textwidth]{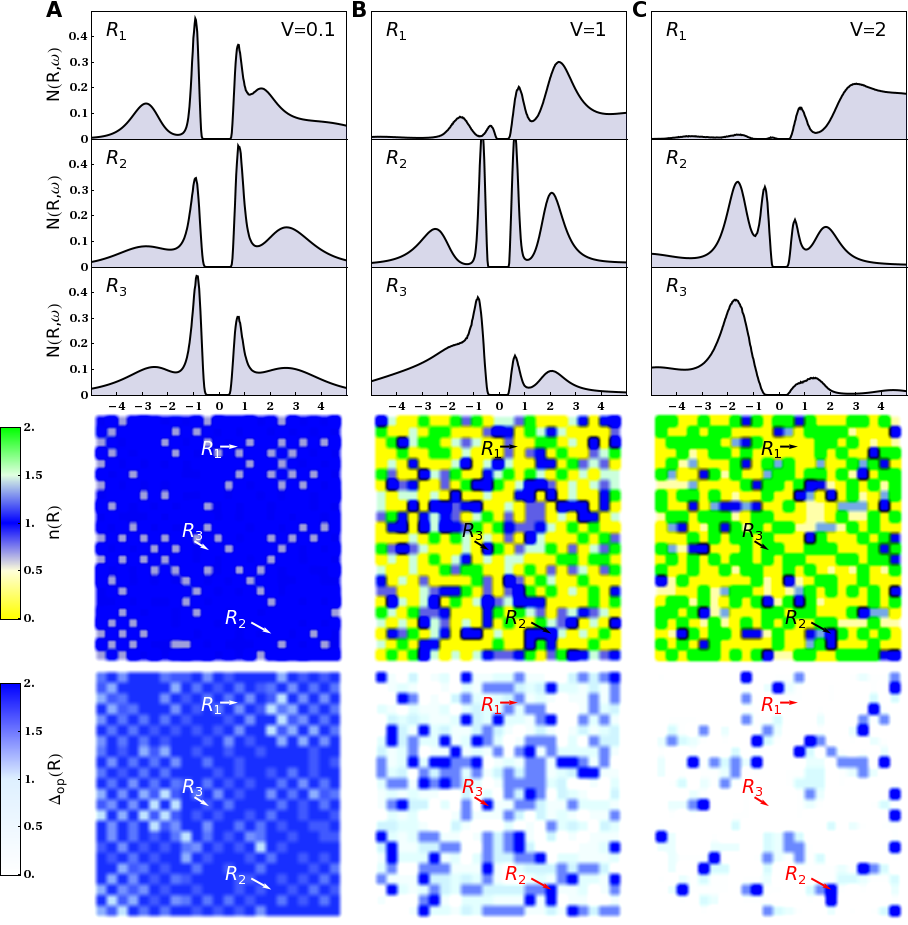}
	\caption{
	Local density of states (LDOS) $N(\RRR,\omega)$, density $n(\RRR)$, and BdG pairing 
	amplitude $\Deltaop(\RRR)$ as a function of disorder strength
	for a montage of nine disorder realizations of $8\times 8$ lattices. Panels A, B, C correspond to $V=0.1, 1, 2$ respectively.
	The LDOS is plotted at three representative sites ${\bf R_i}$. 
	At moderate and strong disorder, 
	site ${\bf R_1}$ is on a high potential hill that is nearly empty, 
	and ${\bf R_3}$ is in a deep valley that is almost doubly occupied. 
	This leads to the characteristic asymmetries 
	in the LDOS in the center and right columns for ${\bf R_1}$ and ${\bf R_3}$. The small
	local pairing amplitude $\Deltaop(\RRR)$ at these two sites is reflected in the absence of coherence peaks in the LDOS.
	In contrast, site ${\bf R_2}$ has a density closer to half-filling,
	leading to a significant local pairing amplitude, a much more symmetrical LDOS, 
	and coherence peaks that persist even at strong disorder.
	\label{localdos}
	}
	\end{figure*}

\myheading{Local probes:}
In Fig.~\ref{localdos} we track the behavior of various local quantities 
with increasing disorder strength $V$.
We show the LDOS $N(\RRR,\omega)$ at representative points,
maps of the spatial variation of the density $n(\RRR)$, and
the BdG pairing amplitude $\Deltaop(\RRR)=\langle c_{\RRR\downarrow} c_{\RRR\uparrow} \rangle$ (which cannot be computed in QMC).
We see that the system becomes increasingly inhomogeneous with increasing disorder, 
as we move from left to right in Fig.~\ref{localdos}.
The SIT occurs due to loss of phase coherence between superconducting islands, seen as
blue patches in the map of the pairing amplitude $\Deltaop(\RRR)$.

We predict experimentally measurable signatures of the local 
density and pairing amplitude in the LDOS $N({\bf R},\omega)$.
Let us focus on three representative sites $\RRR_1$, $\RRR_2$, and $\RRR_3$.
At moderate and strong disorder,
$\RRR_1$ is located on a potential hill, with a low density $n(\RRR_1) \approx 0$ and a negligible pairing amplitude $\Deltaop(\RRR_1) \approx 0$. Thus the
LDOS at $\RRR_1$ is highly asymmetric, with most of the spectral weight at $\omega>0$, for adding an electron. In contrast, $\RRR_3$ is in a potential well, with a high density 
$n(\RRR_3) \approx 2$ and a negligible pairing amplitude $\Deltaop(\RRR_3) \approx 0$.
Thus $\RRR_3$ also has a highly asymmetric LDOS, but most of the spectral weight is at $\omega<0$, for removing an electron.
We believe that MEM correctly captures the gap, coherence peaks, and integrated spectral asymmetry (tested by sum rules); it is much less reliable for high-energy spectral features, which are in any case irrelevant for our purposes.

Finally, $\RRR_2$ lies in a superconducting island close to half-filling, 
$n(\RRR_2) \approx 1$, which permits particle-hole mixing, and therefore a large pairing amplitude $\Deltaop(\RRR_2)$.
The LDOS at $\RRR_2$ is much more symmetrical, with large coherence peaks that persist across the SIT and even in the insulating state.
Note that all the LDOS curves have a clear gap.
We thus find that
symmetrical coherence peaks in the LDOS, and not the local energy gap,
are a clear experimental signature of a local pairing amplitude, which is difficult to probe by other means.
 
\myheading{Discussion:}
We now discuss our results in light of existing theories.
We have ignored the renormalization of the
effective interaction between electrons arising from
changes in screening with increasing disorder~\cite{finkelstein1994}. 
Our point of view is that electronic inhomogeneity   
(that we focus on) is much more important in the vicinity of the SIT than
the disorder dependence of the effective $|U|$ (that we neglect), so long as the latter 
is \emph{not} driven to zero.
This assumption is validated by experiments that find a non-zero gap across the SIT~\cite{sacepe2008}.

Our results are consistent with the absence of a
fermionic or bosonic metal phase in between the superconductor and the insulator. 
Although we have not computed transport here 
(see ref.~\cite{trivedi1996} for an approximate
calculation of the resistivity in the same model), we do not find any
extended low-energy excitations characteristic of a metallic phase. 

The existence of gapped fermions implies a
phase-fluctuation-dominated ``bosonic'' picture for the 
superconductor-insulator transition~\cite{fisher-grinstein-girvin_prl1990,fisher1989}.
However, we must emphasize that we \emph{did not assume} such a bosonic picture 
from the outset. A nontrivial 
aspect of our results is that even though we started with a model of interacting
fermions in a random potential and could have, in principle, obtained (localized) 
gapless fermions in the insulator, we did not find such excitations. The reason
all fermionic excitations are gapped is intimately related to the structure of 
the inhomogeneous local pairing amplitude 
$\Deltaop(\RRR)=\langle c_{\RRR\downarrow} c_{\RRR\uparrow} \rangle$
generated in the presence of large disorder, as we now explain.


	\begin{figure*}[!htb]
	\includegraphics[width=0.9\textwidth]{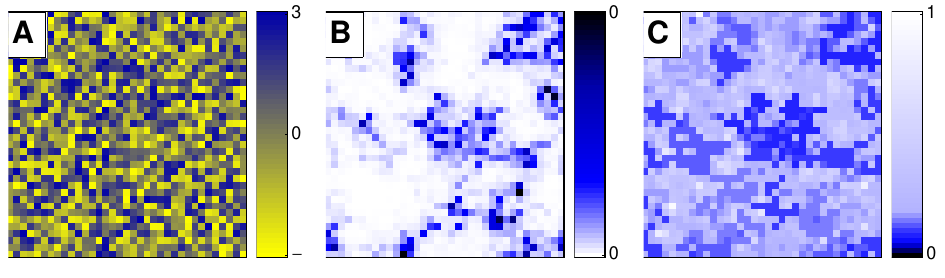}
	\caption{
	\label{puddles}
	\textbf{Emergent granularity:}
	(A) Disorder realization $V(\RRR)$ on a $36\times 36$ lattice at $V=3t$. 
	(B) Local pairing amplitude $\Deltaop (\RRR)$ from a BdG calculation at 
	$\left| U \right| = 1.5t$, $T=0$, and $n=0.875$.  Note the emergent ``granular''
	 structure where the pairing amplitude ``self-organizes'' into superconducting
	 islands on the scale of the coherence length, even though the 
	 ``homogeneous'' disorder potential in (A) varies on the scale
	 of a lattice spacing.
	(C) Local energy gap	$\omegados (\RRR)$ from BdG, 
	defined as the smallest energy at which the local DOS is non-zero ($N(\RRR,\omega) > 0.004$). 
  Note that this gap is finite everywhere and {\it smallest gaps} occur on 
	the SC islands defined by the {\it largest pairing amplitude}.
	}
	\end{figure*}

We show in Fig.~\ref{puddles} that even for ``homogeneous'' disorder, i.e., an uncorrelated random potential $V({\RRR})$ (see panel A),
the pairing amplitude $\Deltaop(\RRR)$ exhibits an
{\em emergent ``granular'' structure} (shown in panel B).
The system self-organizes into superconducting islands, on the scale of the coherence length, with finite  $\Deltaop(\RRR)$, interspersed with insulating regions where $\Deltaop (\RRR)$ is negligible. The spatial variations
of spectral features (asymmetry and coherence peaks) in this inhomogeneous
state were already discussed above in connection with Fig.~\ref{localdos}.

The close connection between inhomogeneity and energy gaps is made clear in
Figs.~\ref{puddles} B and C which demonstrate two striking facts.
We see that (i) there is an energy gap in the LDOS at every site, and 
(ii) small gaps $\omegados(\RRR)$ in the LDOS are spatially
correlated with large $\Deltaop(\RRR)$ SC islands.

A simple way to understand these results is to use the pairing-of-exact-eigenstates approach 
generalized to highly disordered
systems~\cite{ghosal2001}. In the limit of weak attraction, 
pairing gaps out the low-energy density of states in the underlying
Anderson insulator and leads to the islands with non-zero $\Deltaop$
and a small energy gap.
On the other hand, the insulating sea corresponds to
the higher-energy strongly localized states in the system.

From this perspective one can see that the gap
$\omegados$, observed in the spatially average DOS, initially 
decreases with increasing disorder due to a reduction in 
the density of states near the chemical potential in our model.  
(In a real material, the coupling will also decrease~\cite{finkelstein1994}
with disorder).
However, at high disorder, the gap \emph{grows} (consistent with
Fig.~\ref{PhaseDiag}) like $\omegados\approx |U|/(2 \xi_{loc}^2)$ where 
$\xi_{loc}$ is the single-particle localization length~\cite{ghosal2001}.
This is due to the enhanced effective attraction between 
fermions confined to a smaller localization volume $\xi_{loc}^2$.

The phase stiffness (or superfluid density) $\rho_s(T=0)$, on the
other hand, decreases monotonically with disorder as the SC islands 
become smaller and the Josephson coupling between islands becomes weaker.
Thus, even if one starts with a weak-coupling BCS superconductor 
with $\omegados \ll \rho_s$, disorder will necessarily drive it
into the $\omegados \gg \rho_s$ regime.
Eventually quantum phase fluctuations destroy long-range order at $T=0$, 
leading to an insulator whose low-energy excitations are pairs
localized on SC islands.
 
The low-$\rho_s$ regime on the SC side of the SIT leads
to a finite-temperature transition driven by thermal
phase fluctuations ~\cite{emery1995} with $T_c \sim \rho_s(0)$. 
The large energy gap then leads to a marked deviation from conventional BCS theory,
with a pairing pseudogap  in the 
the temperature range $T_c \lesssim T \lesssim \omegados$.
This pseudogap exists even in the weak-coupling regime, provided one is close enough
to the SIT so that $\rho_s \ll \omegados$.

\myheading{Comparison with experiments:}
We describe the connection between our predictions and experiments on the disorder-tuned SIT in systems such as indium oxide, titanium nitride, and niobium nitride films, for which our theory seems to be the most appropriate.  First let us discuss the insulating side of the SIT.
The existence of a gap in the insulator implies activated transport, consistent with early measurements on amorphous InO$_x$ films~\cite{shahar1992}.  In addition, there is evidence for pairs on the insulating side of the transition~\cite{valles_nanohc_science2007} in specially patterned amorphous bismuth films.  

Recent STM experiments are directly relevant to our predictions on the superconducting side of the SIT.
Experiments on homogeneously disordered TiN films~\cite{sacepe2008} have shown that, while $T_c$ goes to zero at the SIT, the STM gap $\omegados$ remains finite, in agreement with Fig.~\ref{PhaseDiag}.  In addition, the gap in the LDOS shows significant inhomogeneity, which supports our picture of emergent granularity (see Figs.~\ref{localdos} and \ref{puddles}).
After our paper was written, we became aware of new experiments that corroborate our predictions.
STM experiments on InO$_x$~\cite{sacepe2011}, TiN~\cite{sacepe_natcomm}, and NbN films~\cite{mondal2011} have all found a pseudogap persisting up to many times $T_c$. 
In particular, they observe a marked suppression of the low-energy DOS together with a destruction of coherence peaks above $T_c$, in complete agreement with our predictions.

We hope that future STM experiments will study in detail the anticorrelation that we predict between the height of the coherence peaks (associated with large pairing amplitude) and the small energy gaps in the local DOS.
The obvious quantum critical scaling between $T_c$ and $\rho_s(0)$ at the SIT, well studied in rather different systems~\cite{hetel2007}, also remains to be tested experimentally in $s$-wave superconducting films.

\myheading{Conclusion:}
In conclusion, we have obtained detailed insights and 
predictions for observable properties of the highly disordered superconducting and
insulating states in 2D films, and of the transition between these states. 
Although we focused on $s$-wave SC films, it has not escaped our attention that  
aspects of our results bear a striking resemblance to the completely different -- and 
much less understood -- problem of the pseudogap in the $d$-wave high $T_c$
superconductors. Features like the loss of low-energy spectral weight persisting across thermal or quantum phase transitions, even as coherence peaks are destroyed, may well be common to all systems where the small superfluid stiffness drives the loss of phase coherence. The 
pseudogap in underdoped cuprates is driven by the proximity to the Mott insulator 
and further complicated by competing order parameters, with disorder
probably playing a secondary role, unlike the disorder-induced pseudogap
near the SIT discussed in this paper.

{\it Correspondence and request for materials shoud be addressed to:} N. Trivedi
(trivedi.15@osu.edu).  
\\
We gratefully acknowledge support from 
NSF DMR-0907275 (KB), 
US Department of Energy, Office of Basic Energy Sciences grant DOE DE-FG02-07ER46423 (NT,YLL),
NSF DMR-0706203 and NSF DMR-1006532 (MR),
and computational support from the Ohio Supercomputing Center. 
KB and YLL performed the numerical calculations; 
MR and NT were responsible for the project planning; 
KB, YLL, MR and NT contributed to the data analysis, discussions and writing.




\end{document}


\title{Supplementary Information for: Single and two-particle energy gaps across the disorder-driven superconductor-insulator transition}\author{Karim Bouadim}
\author{Yen Lee Loh}
\author{Mohit Randeria}
\author{Nandini Trivedi}
\affiliation{Department of Physics, The Ohio State University, Columbus, OH  43210, USA}
\affiliation{}
\begin{abstract}
\end{abstract}
\maketitle

In this supplement we provide details of the determinantal QMC simulations, comparison between QMC and inhomogeneous Bogoliubov-de Gennes (BdG) mean-field theory, and the analytic continuation procedure for extracting real frequency information 
from imaginary time QMC data. 

\myheading{Determinantal QMC:}
We use the determinantal Quantum Monte Carlo (QMC) algorithm~\cite{white1989,blankenbecler} to calculate the 
quantities discussed in the paper, including the imaginary-time 
Green function $G(\RRR;\tau) = - \mean{\calT_\tau  \cccc_{\RRR\sigma} (\tau)  \cdag _{\RRR\sigma} (0)}$ 
and pairing correlation function $P(\tau) = \sum_\RRR \mean{\calT_\tau F (\RRR;\tau) F^\dag (\RRR;0)}$ 
where $F(\RRR,\tau)=  \cccc_{\RRR\dn} (\tau) \cccc_{\RRR\up} (\tau)$.  
We present results for $8\times 8$ square lattices with periodic boundary conditions.
The lattice size is dictated by the need for very accurate QMC data required for analytic continuation.

\begin{figure}[!htb]
		\includegraphics[width=0.94\columnwidth]{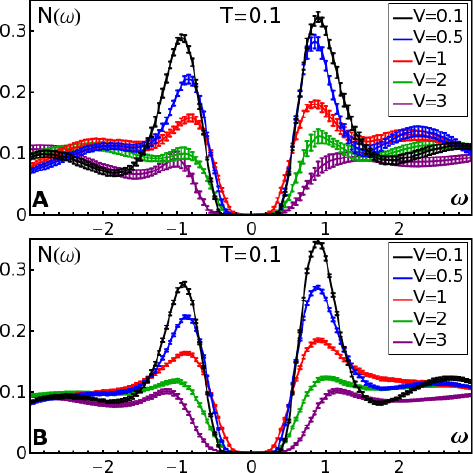}
	\caption{
Comparison of results for the density of states averaged over (A) 10 and (B) 100 disorder realizations. The main features, i.e., hard gaps for all disorder strengths and coherence peaks for $V<V_c$, are robust. Note the reduced statistical fluctuations
in panel B with the increase in the number of disorder realizations. 
	\label{seeds}
	}
	\end{figure}

For a given set of parameters, the simulations are equilibrated for up to $4\times 10^5$ Monte Carlo steps.
The final averages for a single disorder realization are taken over $2\times 10^5$ steps for static quantities and over $4\times 10^6$
for dynamical quantities. We further average over 10 disorder realizations for a given disorder strength. The
resulting maximum absolute errors are $\delta G(\tau)  \sim 10^{-4}$ and $\delta P(\tau) \sim 10^{-2}$.
We have checked the main features in the density of states -- the hard gap at all $V$, and the coherence peaks for $V\lesssim V_c$ -- using extensive simulations with an average over 100 disorder realizations (Fig.~\ref{seeds}).

	\begin{figure*}[!htb]
	\includegraphics[width=1.8\columnwidth]{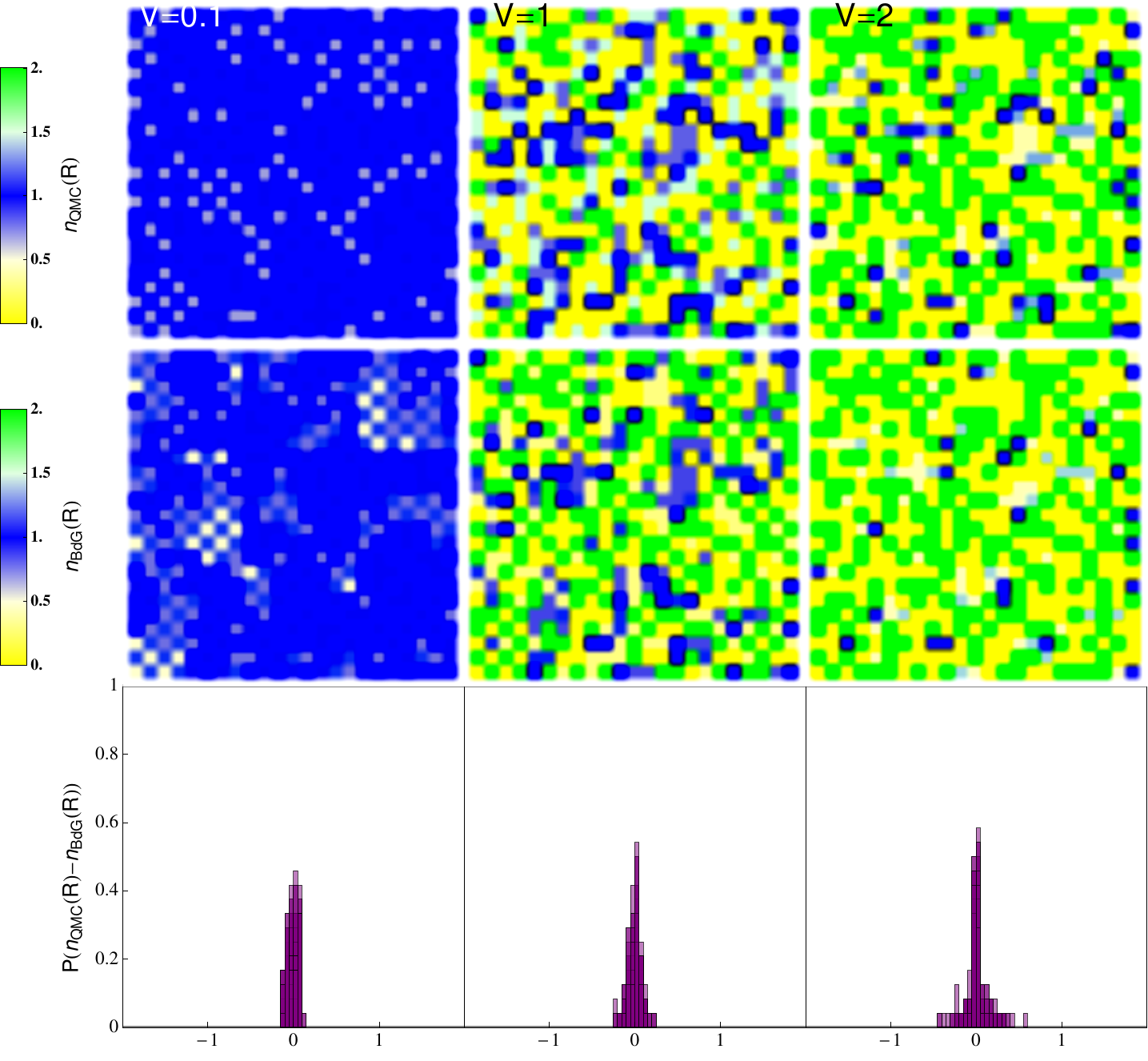}
	\caption{
		\label{RHO_QMC_BdG}
		Comparison of the local density $n(\RRR)$ obtained using QMC (top panels) and self-consistent BdG (middle panels) for one disorder pattern at 
		three different disorder strengths.  
The densities are very similar as seen from the histograms of their differences in the lowest panels, indicating that phase fluctuations have very little effect on $n(\RRR)$.
	}
	\end{figure*}

\myheading{Comparisons of QMC with BdG:}
In Fig.~\ref{RHO_QMC_BdG} we show a comparison of the local density $n(\RRR)$ obtained using QMC and self-consistent 
BdG, including inhomogeneous Hartree shifts, for one disorder pattern at different disorder strengths.  
The close agreement indicates that phase fluctuations, not included in BdG, have very little effect on $n(\RRR)$.
On the other hand, the superfluid stiffness and spectral properties at finite temperatures and large disorder
are greatly affected by thermal and quantum phase fluctuations.

The local density is directly related to  the occupied and unoccupied 
part of the LDOS (see Fig.~4 of the paper) via the sum rules:
$2\int_{-\infty}^{\infty} d \omega f(\omega) N(\RRR,\omega) = n(\RRR)$ 
and $2 \int_{-\infty}^{\infty} d \omega [1-f(\omega)] N(\RRR,\omega) = 2 - n(\RRR)$,
where $f(\omega)$ is the Fermi function and the factor of $2$ comes from spin degeneracy. 
We have tested these sum rules for the
calculated LDOS and find excellent agreement. Further sum rule tests are described
below.

\myheading{Analytic continuation for Green's function:}
We use the maximum entropy method (MEM) to extract the local density of states $N(\RRR,\omega)$
from the imaginary-time Green function $G(\RRR;\tau)$.
The MEM for analytic continuation~\cite{sandvik1998} essentially inverts the Laplace transform
	\begin{equation}
	G(\RRR;\tau) =
	- \int_{-\infty}^{\infty} d \omega \frac{e^{-\tau\omega}}{1 + e^{-\beta\omega}}  N(\RRR,\omega).
	\label{FermionicLaplaceTransform}
\end{equation}
The average DOS $N(\omega)$ is obtained from analytic continuation of $\sum_{\RRR} G(\RRR,\tau)$.
	
\begin{figure*}[!htb]
	\includegraphics[width=\textwidth]{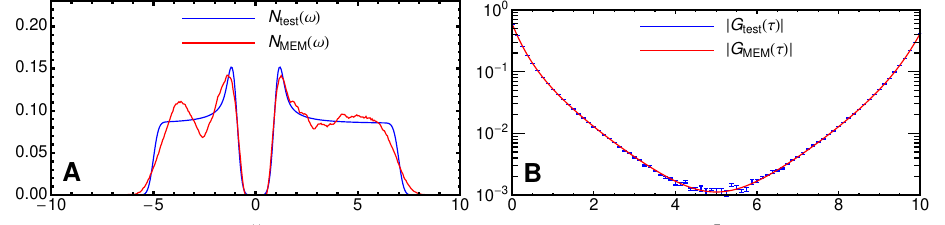}
	\caption{
		Demonstration of analytic continuation using the maximum entropy method (MEM).
		(A)	
		A test spectrum (blue) is chosen, noise is added in the $\tau$-domain,
			and the MEM is used to reconstruct the spectrum (red). The reconstructed spectrum $N_\text{MEM} (\omega)$
			agrees well with the test spectrum $N (\omega)$ at low energies,
			reproducing the correct gap structure,
			although there are deviations at higher energies.
				(B) The reconstructed Green function $G_\text{MEM} (\tau)$ 
			fits the ``data'' $G_\text{data} (\tau)$ to within its error bars.
			The magnitude of the gap in $N (\omega)$ can be estimated 
			by examining the exponential decay constant of $G(\tau)$,
			even without using the MEM.
		\label{mem}
	}
	\end{figure*}

We have performed extensive tests using known model spectra as follows: (i) choose a test spectrum $N(\omega)$; (ii) perform a Laplace transform to obtain the imaginary-time Green function $G(\tau)$; 
(iii) add random numbers $\delta G(\tau)$ drawn independently from a normal distribution of width $\delta G = 10^{-4}$, in order to simulate Monte Carlo statistical error; and finally 
(iv) feed the resulting noisy data, $G_\text{data}(\tau)$, into our MEM routine.  This procedure is illustrated in Fig.~\ref{mem}.
We have concluded that the MEM is adequate for extracting the low-energy features of the spectrum, particularly the gap.

\myheading{Sum rules:}
We have also made extensive sum-rule checks for the spectra obtained from MEM.
We define $M^{(m)}_\RRR$, the $m$th frequency moment of the local density of states at position $\RRR$, as 
	\begin{align}
	M^{(m)}_\RRR
	&= \int_{-\infty}^{\infty} d\omega~ \omega^m N(\RRR,\omega)
	.
	\end{align}
These moments satisfy the following sum rules, which can be derived rigorously by extending the analysis in Ref.~\cite{white1991} to a disordered system:
	\begin{align}
	M^{(0)}_\RRR &= 1,
	\\
	M^{(1)}_\RRR
	&=V_\RRR - \mu + \frac{U}{2} \mean{n_\RRR},
	\\
	M^{(2)}_\RRR - M^{(1)}_\RRR {}^2 
	&= zt^2 + \frac{U^2}{4} \left[ 2 \mean{n_\RRR} - \mean{n_\RRR} {}^2 \right],
	\end{align}
	where $z =4$ is the coordination number.
Differentiating Eq.~\eqref{FermionicLaplaceTransform} shows that the moments are also related to the values and derivatives of the Green function at $\tau=0$ and $\tau=\beta$,
	\begin{align}
	M^{(m)}_\RRR &= (-1)^m \left[  
		\frac{\dd^m G_\RRR}{\dd \tau^m} (0) 
	+	\frac{\dd^m G_\RRR}{\dd \tau^m} (\beta)
	\right]
	.
	\end{align}
QMC simulations produce very accurate results for $G_\RRR(0)$ and $G_\RRR(\beta)$, with absolute errors of about $10^{-5}$.  
The MEM analytic continuation procedure fits these data points to within the error bars.
We have verified that the MEM LDOS satisfies the moment sum rules with a fractional error of less than $0.001\%$ for $M^{(0)}$, less than $1\%$ for $M^{(1)}$, and about $1\%$ for $M^{(2)}$.
In contrast, the LDOS obtained from BdG calculations is found to violate the sum rules.  

\myheading{Analytic continuation for pairing correlation:}
The pair spectrum $P''(\omega)$ is related to the pairing correlation function $P(\tau)$
via 
\begin{equation}
P(\tau) = \int_{-\infty}^{\infty} \frac{d \omega}{\pi}
\frac{e^{-\tau\omega}}{1 - e^{-\beta\omega}}  P''(\omega).
\label{BosonicLaplaceTransform}
\end{equation}
We write this in a manner similar to the fermionic result of Eq.~(\ref{FermionicLaplaceTransform}) as
\begin{equation}
P(\tau) = \int_{-\infty}^{\infty} \frac{d \omega}{\pi}
\frac{e^{-\tau\omega}}{1 + e^{-\beta\omega}}  P''(\omega)  \coth \frac{\beta\omega}{2}.
\label{BosonicLaplaceTransform2}
\end{equation}
and use MEM to invert the Laplace transform.

	\begin{figure*}
	\includegraphics[width=\textwidth]{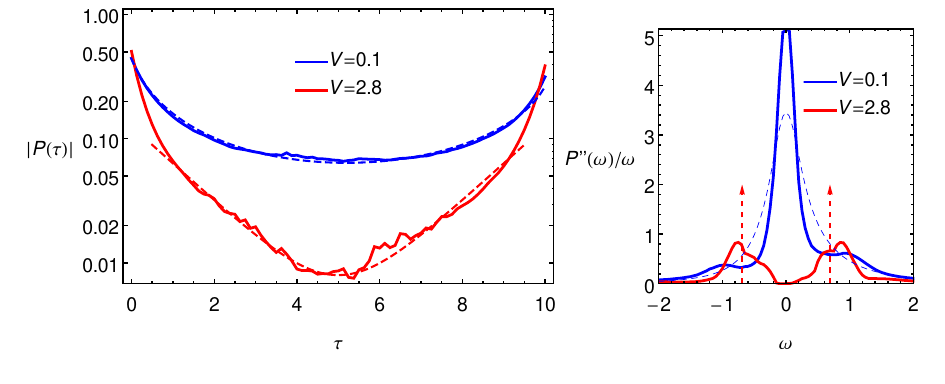}
	\caption{
		(Left)
		Two-particle imaginary-time Green functions from QMC (each averaged over 10 disorder realizations).
		For strong disorder (solid red curve) the curves contain straight-line sections, i.e., regions of exponential behavior,
			suggesting a gapped spectrum,
		whereas for weak disorder there is no clear straight-line behavior.
		Dashed curves correspond to model spectra in right panel.
		(Right)
		Solid curves are two-particle spectra extracted using maximum entropy method.
		The dashed blue curve is a gapless, Lorentzian-like spectrum.
		The dashed red curve is a gapped spectrum.
		\label{TwoParticleMEMDetails}
	}
	\end{figure*}

We have made extensive checks on the MEM results for $P''(\omega)$. 
As an example, we illustrate in Fig.~\ref{TwoParticleMEMDetails} 
how very simple analytic expressions fit to the $P(\tau)$ QMC data can give us
confidence in the ability of MEM to discriminate between gapped and 
gapless spectra. We show in the left panel of 
Fig.~\ref{TwoParticleMEMDetails} the
two-particle Green function $P(\tau)$ at two disorder strengths $V=0.1$ 
(superconducting state) and $V=2.8$ (insulating state).
The QMC data are at low temperature $T=0.1$ and 
averaged over 10 disorder realizations in each case, with  
error bars suppressed for clarity.
At strong disorder the QMC data can be fit by an 
exponential decay in imaginary time
(with periodicity $\beta$). This leads to
a spectrum with two delta functions 
${P''(\omega)}/{\omega} \propto \delta(\omega-\Omega) + \delta(\omega+\Omega)$ 
with $\Omega=0.69$.
At weak disorder the data can be fit better with a gapless, 
Lorentzian-like spectrum
	\begin{align}
	\frac{P''(\omega)}{\omega} &\propto
		 \frac{1}{\omega} \tanh \frac{\beta\omega}{2} 
		 \times
		 \begin{cases}
		 \frac{1}{1+\omega^2/\Omega_+^2} & \omega>0 \\
		 \frac{1}{1+\omega^2/\Omega_-^2} & \omega<0. 
		 \end{cases}
	\end{align}
The fact that $\Omega_+=1.4$ and $\Omega_-=0.8$ in this simple fit
shows the particle-hole asymmetry evident in $P(\tau)$.
We emphasize that in both cases, these simple functional forms capture
the essential characteristics -- gapped vs. gapless -- of the 
MEM spectrum as illustrated in the right panel of 
Fig.~\ref{TwoParticleMEMDetails}. 

	\begin{figure*}
	\includegraphics[width=\textwidth]{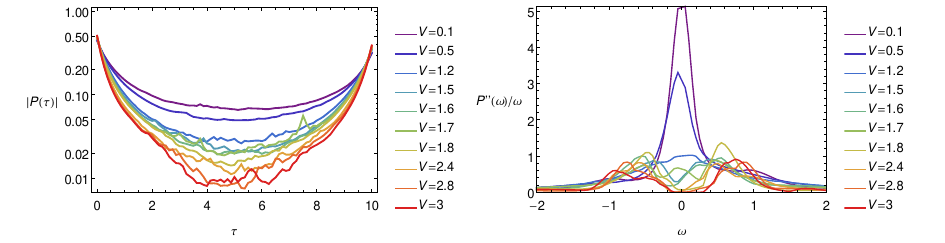}
	\caption{
		(Left)
		Two-particle imaginary time Green function from QMC
			for various disorder strengths (averaged over 10 seeds).		
				(Right)
		Corresponding spectra from MEM.
		\label{TwoParticleMEMManyCurves}
	}
	\end{figure*}
	
We show in greater detail in Figure~\ref{TwoParticleMEMManyCurves} how $P(\tau)$ and the corresponding MEM spectra $P''(\omega)/\omega$ evolve with increasing disorder 
strengths. These data are averaged over 10 disorder realizations in each case.
As remarked in the main text, the main trend is qualitatively clear:
$P''(\omega)/\omega$ has a single peak in the superfluid state,
which splits into two peaks in the insulating state ($V\gtrsim 1.6$)
that separate with increasing disorder.


